\documentclass[journal]{IEEEtran}
\usepackage{threeparttable}
\usepackage[linesnumbered,ruled]{algorithm2e}
\usepackage{bm}
\usepackage{graphicx,hyperref}
\usepackage{multirow,booktabs}
\usepackage{amsmath,subfigure,amssymb }

\ifCLASSINFOpdf
\else
\fi
\begin{document}
%
\title{Multi-Source Interactive Resilient Fusion Algorithm Based on RIEKF}
%
%
%

\author{Ye Xiaoyu, Song Fujun, Zhu Xiaohu and Zeng Qinghua
\thanks{The authors are with the School of Aeronautics and Astronautics, Sun Yat-sen University, Shenzhen 518107, China. e-mail: \href{mailto:yexy39@mail2.sysu.edu.cn}{yexy39@mail2.sysu.edu.cn}.}}
\maketitle

\begin{abstract}
As the number of heterogeneous redundant sensors on unmanned aerial vehicle
(UAV) increases, onboard sensors require a more rational and efficient credibility evaluation system and a resilient fusion framework to achieve the essence of seamless sensor group switching. A simple and efficient sensor credibility evaluation system is proposed to guide the selection of the optimal multi-source sensor submodel combination, thereby providing key model prior knowledge for multi-source resilient fusion. Furthermore, a multi-model interactive resilient fusion framework based on RIEKF is proposed, utilizing the defined sensor credibility indexes to guide the design of the model transition probability matrix, thereby reducing the sensitivity of submodel weights to fusion stability and solving the problem of the model transition matrix lacking a basis for adjustment. Model weights are updated in real time through credibility prior information and submodel posterior probabilities, thus leveraging the adaptive resilience advantage between models to achieve seamless switching between submodels in complex environments. Experimental results show that the algorithm presented in this paper, without using any sensor fault diagnosis and isolation logic, without setting any complex detection timing and thresholds, demonstrates a resilience advantage, thereby enhancing the adaptability of the state estimation system in complex environments.
\end{abstract}

\begin{IEEEkeywords}
resilient fusion, multisensors fusion, invariant Kalman filter, state estimation, GNSS denied environment.
\end{IEEEkeywords}

%
\IEEEpeerreviewmaketitle

\section{Introduction}
%
%
%
%
\IEEEPARstart{W}{ith} the rapid development of unmanned systems, navigation state estimation systems have become a crucial means for autonomous systems to perceive their own state and environment \cite{RN2411,RN2027}. 

In recent years, the development of airborne sensor suites towards multi-source, heterogeneous, and redundant configurations has accelerated. Traditional sensor fusion frameworks have focused more on single combinations, enhancing adaptability to the environment from a performance perspective but often overlooking the coordination between sensors. Different heterogeneous sensors each have unique applicable scenarios and functions. Due to the strong heterogeneity in the functionalities and error characteristics of different sensors, this poses significant challenges to designing fusion algorithms based on traditional approaches. Typically, a complete fusion algorithm needs to include effectiveness detection, fault discrimination, diagnosis, and isolation for each sensor. Representative technologies include Fault Detection and Diagnosis (FDE) and autonomous integrity monitoring technology (AIM), with notable work by Jurado \cite{jurado2019} proposing the Autonomous and Resilient Management of All-source Sensors (ARMAS) \cite{gipson2022resilience}, which divides the resilient fusion of multi-source sensors into four processes: (1) fault diagnosis and isolation, (2) validity determination, (3) calibration, and (4) model reconstruction. The complexity of these processes demands high requirements for detection and diagnostic logic \cite{meng2022resilient}, significantly increasing the design and debugging difficulty of algorithms.

In engineering, it is common to use a large number of if-else statements to handle sensor validity checks, such as with the PX4 ECL toolbox \cite{ECL,APM}, by utilizing defined error tolerance thresholds to determine if a sensor is suitable for use at any given moment. Relying on threshold-based methods often leads to false alarms and missed detections. This direct method becomes extremely complex as the number of heterogeneous sensors increases. Therefore, there is an urgent need for a multi-source resilient fusion mechanism. The fundamental pursuit is to maximize the residual value of sensors, select the optimal combination based on the health of the sensors, and reduce the usage rate of if-else statements, much like human judgment based on scenarios, in establishing future navigation systems that are efficient, reliable, resilient, and sustainable.

Research on the credibility evaluation system for multi-source integrated navigation systems remains notably limited. Dong et al. \cite{DongEvaluation2022} reviewed the progress in performance evaluation methods for inertial navigation systems, pointing out issues such as the incompleteness of the index system and the lack of precision in weightings. Cheng et al. \cite{cheng2019ow} divided the index system of the INS/GNSS integrated navigation system into functional layers, index layers, and device layers, providing a detailed classification. However, the strong correlation between different indexes often makes it difficult to assess individual indexes, leading to extreme vagueness and roughness, making it challenging to apply directly in engineering practice. Some sensors can directly output reference indexes on the device to assist in determining the current performance state of the sensor, such as GPS receivers that can output the DOP (Dilution of Precision) index to assess the quality of positioning signals \cite{karaim2018gnss}, TOF (Time of flight) rangefinders that can output the signal strength obtained from laser reflection for judgment, and Lidars that can evaluate using the Radar Cross Section (RCS) index, etc. However, some sensors require the use of kinematic models to calculate navigation information, such as barometers, magnetometers, and visual cameras. Barometers convert sensitive atmospheric pressure into barometric altitude; visual odometry's pose calculation is completed by the PnP (Perspective-n-Point) process, where the number and quality of feature point recognition significantly affect pose estimation accuracy \cite{Likailin2023}. Therefore, there is a lack of a unified evaluation system for the credibility of multi-source sensors. Alessandro et al. \cite{fornasier2021vinseval} established a unified evaluation framework for VINS (Visual-Inertial Navigation System) to conduct repeatability analyses of different VINS methods in specific environments and sensor parameters. Won et al. \cite{won2013geometrical,won2014selective} proposed using weighted DOP as an accuracy index for position and attitude in vision-based navigation, guiding the adaptive fusion weights of the INS/GNSS/VISION model. Chiu et al. \cite{chiu2014constrained} innovatively proposed a constrained optimization mechanism for selecting the optimal subset of sensors, using different sensors' observability \cite{sun2008observability}, expected accuracy, etc., as evaluation indexes. However, this evaluation strategy relies more on prior sensor information, which is challenging to use in complex environments to judge sensor performance solely based on this information, as it is constantly changing. In this context, the choice of the optimal sensor combination also needs to integrate more post-information from model feedback, such as model states, to assist in judgment, enhancing adaptability to the environment.

Secondly, the need for an intelligent sensor group selection mechanism is not only to choose the optimal subset of sensors but also to give a trust priority to the remaining sensors, incorporating it as a key index into the state estimation system to realize the connotation of "resilient" fusion. That is to deeply mine the complementary information of sensor data to improve the robustness and environmental adaptability of the navigation system, rather than merely discarding data directly. Currently, the Interactive Multi-Model (IMM) approach is one potential solution for resilient fusion, which has been extensively researched and applied in the field of target tracking. The primary feature of IMM is its ability to "soft switch" between different models, adapting to the uncertainties between discrete and continuous states, making it highly suitable for tracking maneuvering targets with discrete motion uncertainties and continuous motion. In recent years, IMM has also been applied in multi-source information fusion systems. Jo et al. \cite{jo2011interacting} proposed an intelligent vehicle navigation algorithm based on IMM, which models different system dynamics based on the EKF framework and assumes that the system will follow one of a finite number of different dynamics models during operation. It adaptively adjusts the corresponding model weights by combining system state information for update iterations. This adaptive capability gives IMM estimators advantages over direct method estimators, such as Kalman filters. Min et al. \cite{min2019kinematic} also applied IMM in vehicle navigation, integrating vision/INS/GPS and two types of vehicle dynamics models. The former studies are mostly based on different motion models, not specifically for redundant navigation systems. Meng et al. \cite{meng2022resilient} proposed a resilient interactive sensor-independent-update (ISIU) algorithm based on IMM theory. This algorithm changes the system prediction model to three measurement subsystems: RTK/INS, LiDAR/INS, VINS/INS, and uses the posterior probability as the basis for calculating dynamic sensor subsystem weights, eliminating the need for additional complex fault-tolerant fault diagnosis strategies.

The main limitation of the ISIU strategy based on IMM is its fusion accuracy, which is highly sensitive to the model transition probabilities \(\pi\), i.e., the Markov chain. The model transition probabilities, as prior values, indicate the trust priority among different observed sensor combinations. The current practice, as described in the literature \cite{meng2022resilient}, is to manually tune the size of \(\pi\). Inappropriate transition probabilities can significantly impact the fusion results. Therefore, transition probabilities need to be determined in a more scientific manner. Moreover, the transition probability \(\pi\) is not a constant matrix; the transition probabilities between different sub-filter models in a multi-source fusion system need to be adaptively adjusted as the system state changes. It is necessary to guide the design of the filter transition probability matrix based on the current credibility of heterogeneous sensors. The main contributions of this paper are as follows:

(1) Propose a simple, dimensionless sensor credibility evaluation system, selecting the optimal and sub-optimal multi-source sensor sub-model combination architectures based on the credibility of different heterogeneous sensors, providing model prior information for multi-source resilient fusion.

(2) Propose an improved multi-source resilient fusion algorithm using invariant Kalman filtering as the sub-model, utilizing the sensor credibility index to design the multi-model transition probability matrix, thereby reducing the sensitivity of the sub-model weights.

(3) Validate the proposed algorithm with real flight data on a UAV redundancy IMU/GPS platform, and inject multiple faults such as heavy-tailed noise and communication denial to fully test the algorithm's resilient advantages.

The remainder of this document is structured as follows: Section \ref{sensor_reliability_index} delineates the definition of sensor reliability indices, while Section \ref{sensor_select} elucidates the methodology for optimal sensor selection. Section \ref{IMM-based Multi-Source Resilient Fusion} elaborates on the proposed resilient fusion framework grounded in the IMM pipeline. Section \ref{exp_on_multi_GPS/INS} presents experiment validation of the algorithm's robustness and its resilience advantages under various sensor failure scenarios, conducted through experiments on redundant flight control systems of UAVs. Section \ref{Discussion} focuses on a critical discourse concerning the limitations encountered during the application of the proposed algorithmic framework, along with a prospective outlook. Section \ref{Conclusion} provides a conclusion for the entire paper.
\section{Definition of Sensors Credibility Index}\label{sensor_reliability_index}
Academician Wang Wei pointed out that the credibility of multi-source autonomous navigation systems is an intrinsic attribute describing the reliability of the functions and results of multi-source autonomous navigation systems. It measures the ability to trust the calculated results after various information sources have undergone interference, fault detection, fault identification, fault elimination, and system reconstruction \cite{wangwei2023}. Aiming to provide pose and velocity state estimation, a multi-source navigation system integrates various heterogeneous sensors, complementing each other's advantages and providing quantifiable constraints for the system state. Table \ref{tab:sensors} shows the classification of commonly used onboard sensors for UAVs and their association with pose states.

\begin{table*}
    \caption{Classification of heterogeneous sensors and their association with poses\tnote{a}}
    \label{tab:sensors}
    \small
    \centering
    \begin{threeparttable}[b]
    \begin{tabular}{ccccccccc}
    \toprule
        \textbf{Sensor} & \textbf{Signal Type} & \textbf{Measurement} & $\boldsymbol{\theta}/ \boldsymbol{\gamma}$ & $\boldsymbol{\phi}$ & $\boldsymbol{v}_{x,y}$ & $\boldsymbol{v}_z$ & $\boldsymbol{p}_{x,y}$ & $\boldsymbol{p}_{z}$\\
        \midrule
         IMU & Relative Signal  & Rotation, Acceleration & 1 & 0 & 1 & 1 & 1 & 1\\
         GPS & Global Signal  & Velocity, Position & 1 & 1 & 1 & 1 & 1 & 1\\
         RTK & Global Signal & Velocity, Position & 1 & 1 & 1 & 1 & 1 & 1\\
         Magnetometer & Global Signal & Heading & 0 & 1 & 0 & 0 & 0 & 0\\
         Barometer & Global Signal & Altitude & 0 & 0 & 0 & 1 & 0 & 1\\
         Laser Altimeter & Global Signal & Altitude & 0 & 0 & 0 & 1 & 0 & 1\\
         Mono (Stereo) Camera & Relative Signal & Visual Features & 1 & 0 & 1 & 1 & 1 & 1\\
         3D Lidar & Relative Signal & Rotation, Displacement & 1 & 0 & 1 & 1 & 1 & 1\\
         mmWave Radar & Relative Signal & Rotation, Displacement & 1 & 0 & 1 & 1 & 1 & 1\\
         \bottomrule
    \end{tabular}
    \begin{tablenotes}
    \item[a] Note: The 0 in the IMU heading (\(\phi\)) column does not mean that IMU cannot compute heading information, but rather that low-cost IMUs cannot align to the heading, failing to provide a valid heading reference.
    \end{tablenotes}
    \end{threeparttable}
\end{table*}

Due to the heterogeneous characteristics of sensors, there is currently no mature and unified standard or model to evaluate and judge the credibility and performance boundaries of sensors. In this paper, based on the basic connotation of sensor credibility, we categorize it into two aspects: one is the expected performance of the sensor, which shows the expected accuracy (\(3\sigma\)), that is, the performance boundary that can be achieved under normal operating conditions. The larger the value, the higher the expected accuracy of the sensor, with a lower limit but no upper limit. The second is the realtime performance of the sensor in specific navigation environments, which represents the degree of performance release during the sensor's operation. Since different sensors have their applicable range boundaries, the performance of sensors will be affected in multiple aspects outside these boundaries. Therefore, when considering sensor credibility, it is necessary to comprehensively consider the above two indexes. Both credibility performance indexes are indispensable, which is also an important factor considered in our selection of the optimal sensor combination.

The sensor credibility index is comprehensively defined as the product of two factors to better represent the interaction or compound effect between them. These two indexes influence each other and have a zero effect, meaning that if the realtime credibility of a sensor is zero, no matter how high the expected credibility is, the final performance index will be zero, which aligns more closely with real-world scenarios. Therefore, the comprehensive definition of the sensor credibility index is as shown in Eq.~( \ref{ch6:eq:index}):
\begin{equation}\label{ch6:eq:index}
    H = E \times P
\end{equation}
where \(E\) represents the expected credibility index, and \(P\) represents the realtime credibility index. The two indexes are elaborated on below.
\subsection{Expected Credibility Index}
For each sensor, given the heterogeneity of the indexes, it is necessary to find a universal way to describe them. Considering that most multi-source navigation systems are primarily based on IMUs, that is, they use multi-source fusion to constrain and compensate for the INS, taking advantage of the IMU's high refresh rate, completeness, and short-term high performance. However, the error accumulation characteristic of INS requires the assistance of other sensors to correct errors through auxiliary correction methods. Therefore, the error accuracy of multi-source sensors will directly affect the navigation accuracy of the INS main system.

Besides the IMU, the errors of other sensors are closely related to one or more of the three aspects: attitude \(\boldsymbol{\phi}\), velocity \(\boldsymbol{v}\), and position error \(\boldsymbol{p}\). For example, GPS directly affects the accuracy of velocity and position measurements, vision sensors are closely related to pose measurements, and magnetometers can directly provide heading angle information, among others. When comprehensively evaluating the expected performance of heterogeneous sensors, it is necessary to find a mechanism that can balance the weights of these three error indexes well, to universally describe the expected performance of the sensors.

Using the short-term inertial navigation error characteristics, which can effectively separate the impact of position errors caused by different error sources, this paper describes the expected performance of sensors based on these short-term inertial navigation error properties \cite{RN2017}, as shown in Eq.~(\ref{ch6:eq:sensor_err_cal}).
\begin{equation}\label{ch6:eq:sensor_err_cal}
\begin{gathered}
    e_{b_g} = g^n \cdot t^3 \cdot \delta b_g /6,
    \, \,e_{b_a} = \delta b_a \cdot t^2/2,\\
    e_{\phi_{xy}} = g_0 \delta   \phi \cdot t^2 / 2,\, \,
    e_{\psi} = \delta \psi \triangle v \cdot t,\\
    e_{v} = \delta v \cdot t, \, \,
    e_{p} = \delta p,\\
    e_{total} = e_{\phi_{xy}} + e_{\psi} + e_{v} + e_{p}\\
\end{gathered}
\end{equation}
where, \(e_{\phi_{xy}}, e_{\psi}, e_{v}, e_{p}\) respectively represent the positioning recursive error of the inertial navigation within the period \(t\) caused by horizontal attitude error \(\delta \phi\), heading error \(\delta \psi\), velocity error \(\delta v\), and position error \(\delta p\). \(e_{total}\) is the total sum of positioning calculation errors. Considering that in vehicles, positioning information is a critical source of information, using the total sum of position errors to unify dimensions can eliminate inconsistencies due to sensor heterogeneity in the evaluation of the sensor credibility system.

Since the positional error is decreasing relative to the sensor's accuracy, it is necessary to take the reciprocal of the error value when describing the expected reliability of the sensor:
\begin{equation}
        E = \frac{1}{e_{total}}
\end{equation}
Based on this, an index for the expected reliability of the sensor can be established.

\subsection{Realtime Credibility Index}
The realtime reliability index represents the performance of the sensor system during live operation, described by \(0-\rho_{max}\). A smaller value indicates poorer current sensor performance, a larger value indicates that the current sensor performance is closer to the performance boundary, and a value of 0 indicates that the sensor is experiencing irreconcilable failure or is not operational, with \(\rho_{max}\) being the upper bound of realtime reliability.

Determining the realtime reliability of the sensor becomes a critical issue. It should be noted that some sensors can directly output reference indexes on the device to assist in determining the current sensor performance status. For example, GPS can evaluate the quality of positioning signals by obtaining DOP parameters, visual SLAM can indirectly assess the reliability of pose estimation by the number of feature points, and laser rangefinders can evaluate signal accuracy by the output strength of the reflected signal. However, other sensors require the assistance of covariance indexes within the filtering model to determine, as shown in the Eq. (\ref{cal_covariance}).
\begin{equation}\label{cal_covariance}
\begin{aligned}
    \delta \bm{x} &= \bm{z} - \bm{H}  \bm{x}\\
    \operatorname{Cov}[\delta \bm{x}] & =E\left[\delta \bm{x} \delta \bm{x}^T\right] \\
    \bm{S} & =\bm{H}\bm{P}\bm{H}^{\top}+\bm{R}
\end{aligned}
\end{equation}
here, \(\bm{R}\) is the measurement covariance matrix, \(\bm{H}\) is referred to as the observation matrix, \(\delta \bm{x}\) represents the measurement innovation, and \(\bm{S} = \operatorname{Cov}[\delta \bm{x}]\) is the covariance matrix of measurement innovation, providing information on the error distribution. The sensor's positioning reliability level is represented by \(DOP = \sqrt{\text{tr}(\bm{S})}\), where \(\text{tr}\) is the trace of the matrix. If the diagonal elements of the covariance matrix are relatively stable over different times or conditions, this indicates that the sensor maintains good consistency in different environments. The smaller the DOP value, the higher the precision. Therefore, the realtime reliability of the sensor determined with the aid of the covariance index within the model is the reciprocal of the DOP value:
\begin{equation}
    P = \frac{1}{DOP} = \frac{1}{\sqrt{\text{tr}(\bm{S})}}
\end{equation}

For the evaluation of multi-source sensor performance indexes, dimensionallessness is an indispensable operation in the assessment to accurately eliminate the incommensurability of indexes \cite{dong2021thesis}. Cheng et al. \cite{cheng2019ow} proposed a dimensionless method that combines the extremum method with the exponential function method, but the index function curve of this algorithm is relatively fixed and lacks universality. The curve of realtime performance evaluation indexes for sensors is a highly nonlinear process, requiring dynamic adjustment of weight distribution according to specific sensor situations. In this paper, combining Eq.~(\ref{dimensionless_eq}), the parameters of the dimensionless function are iteratively optimized by selecting feature points and fitting the curve. Eq.~(\ref{dimensionless_eq}) defines the curve fitting function for the dimensionless function of the universal sensor realtime performance index, divided into two types of curves. One is an increasing curve, where the realtime performance index increases as the sensor index increases, such as in visual sensors, where the algorithm's credibility increases with the number of feature points under normal operating conditions. The other is a decreasing curve, where the realtime performance index decreases as the sensor index increases, such as the DOP value of GPS, where a larger value indicates lower algorithm credibility.
\begin{figure*}[!t]
\begin{equation}\label{dimensionless_eq}
\begin{aligned}
    & f_{rise}\left(x_i\right)= \begin{cases}10, & x_i \leq \min x_i \\
    a \cdot\left(\frac{\max x_i-x_i}{\max x_i-\min x_i}\right) e^{ b \cdot\left(\frac{\max x_i-x_i}{\max x_i-\min x_i}\right)}, & \min x_i<x_i<\max x_i \\
    0, & x_i \geq \max x_i,\end{cases} \\
    & f_{down}\left(x_i\right)= \begin{cases}10, & x_i \geq \max x_i \\
    a \cdot\left(\frac{x_i-\min x_i}{\max x_i-\min x_i}\right) e^{b \cdot\left(\frac{x_i-\min x_i}{\max x_i-\min x_i}\right)}, & \min x_i<x_i<\max x_i \\
    0, & x_i \leq \min x_i .\end{cases}
\end{aligned}
\end{equation}
\end{figure*}
here, \(x_i\) represents the unique index of the \(i\)th sensor, \(\max x_i\) and \(\min x_i\) respectively represent the maximum and minimum values of the defined performance indexes, \(a\) and \(b\) are the fitting coefficients.

\begin{figure}[!t]
    \centering
    \begin{minipage}{0.9\linewidth}
		\centering
		\includegraphics[width=0.9\linewidth]{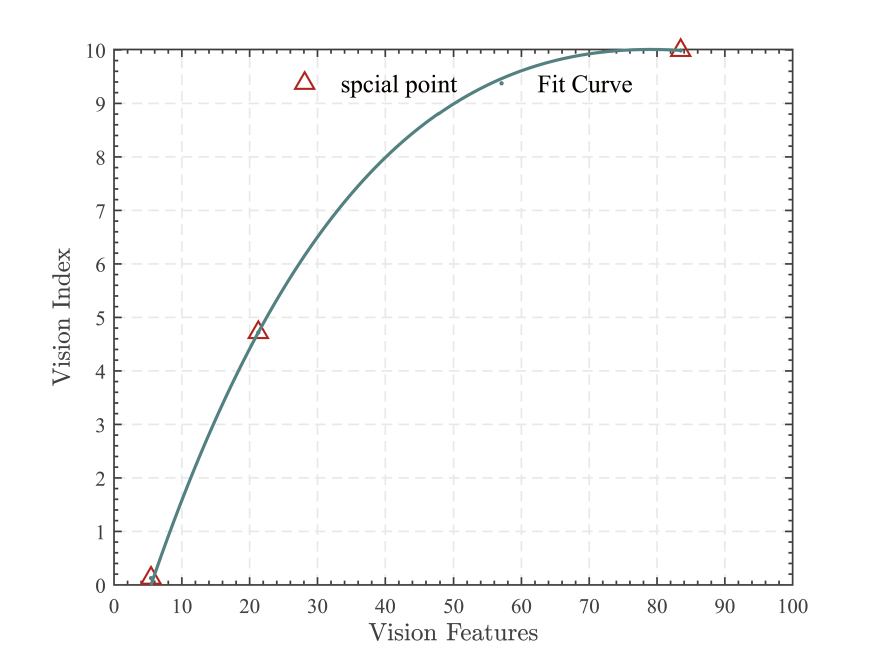}
		\caption{Within the operational environment of the algorithm, the credibility of the realtime performance index for camera sensors escalates with the increment in the number of feature points.}
		\label{vision}
	\end{minipage}
 \qquad
\begin{minipage}{0.9\linewidth}
		\centering
		\includegraphics[width=0.9\linewidth]{figures/vision_index.pdf}
		\caption{The usability of GPS sensors diminishes as the DOP metric increases.}
		\label{GPS}
	\end{minipage}
\end{figure}

Fig. \ref{vision} and Fig. \ref{GPS} shows the curve of the dimensionless function for realtime performance indexes designed for visual and GPS sensors. The triangles in the figure mark the feature points that are more in line with the actual sensor situations. Based on Eq.~(\ref{dimensionless_eq}), the autonomously fitted function curve can conveniently and accurately dimensionless the unique performance indexes of sensors to eliminate differences.
\section{Sensor Subsets Selection}\label{sensor_select}
\begin{figure*}[!t]
    \centering
    \includegraphics[width=0.8\textwidth]{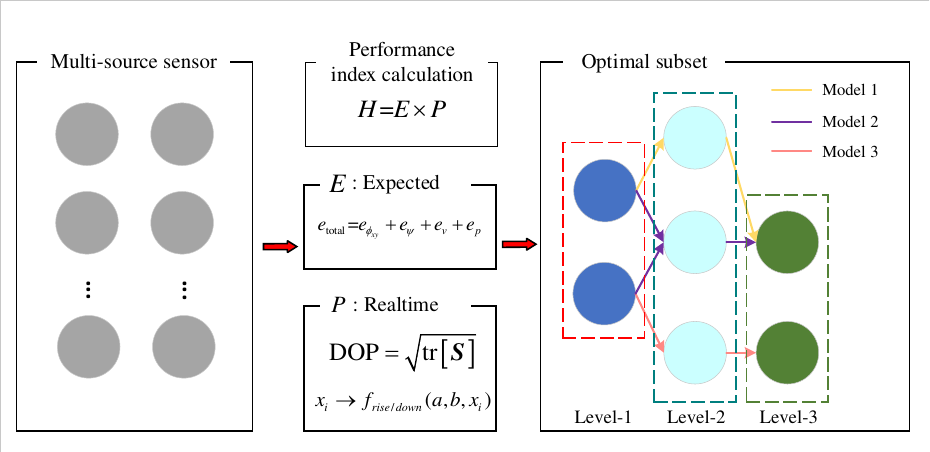}
    \caption{Sensor subset selection framework. The optimal subset is primarily divided into three levels.}
    \label{sensor_subset}
\end{figure*}

In this section, we propose an optimal subset classification method for sensors based on credibility indexes and state coverage. As shown in Fig.~(\ref{sensor_subset}), the selection of sensor subsets needs to fully consider the coverage rate of sensors on navigation parameters and credibility indexes, capable of providing a satisfactory hierarchical sensor subset fusion scheme under limited resources.

The optimal subset is primarily divided into three levels. The first level consists of IMUs, due to their high autonomy and their capability to independently provide pose estimation for the system, which gives them a higher priority over other sensors, making them the core sensors in many state estimation systems. The second level includes sensors that can constrain multiple state variables simultaneously, playing a major role in error correction for state estimation systems primarily based on INS. The third level comprises sensors that only impose constraints on specific, fixed states of the system, usually with a lower degree of coupling, and are used to constrain key states of the system in certain special environments.

For the first level, if there are multiple redundant IMUs in the system, we rank them according to the expected performance indexes of the IMUs, thereby selecting the most reliable IMU for fusion, with the rest serving as redundant backups. Experiments in literature \cite{zhang2020lightweight} have verified that fusing multiple IMU arrays can enhance positioning accuracy, but the improvement is not significant; hence, only the best-performing IMU is chosen for fusion.

In the selection of the second and third layer sensor groups, the credibility indexes of the current multi-source sensor groups are calculated in realtime and ranked. Additionally, a tree structure needs to be formed with the first-level IMU nodes. The selection criteria for the tree structure are to ensure the optimal and sub-optimal combinations of sensor credibility indexes under the condition that the system states are all constrainable, as seen in Table \ref{tab:sensors}. For the six navigation state sets, if a sensor has a constraining effect, it is marked as 1; if it is unobservable or weakly observable for that state, it is marked as 0. The highest-precision IMU selects the currently most credible sensor and assesses the coverage of the indexes. If there are states not fully covered, the search continues for the most credible sensor that can cover the current state variable as a level-3 supplement. If sensors in L3 have better credibility than those selected in L2, then among these, a sensor with the optimal performance index for different covered state variables is selected and added to the chosen tree structure.

\begin{algorithm}[!t]
\small
\caption{Sensor Optimization Algorithm}
    \KwIn{Sensors $S_i$, Sensor Credibility Indices $C_{S_i}$, and State Coverage Matrix $M_{S_i}$}
    \KwOut{Fusion order and framework for Level-2 and Level-3 sensors $\mathcal{T}_1, \mathcal{T}_2, ..., \mathcal{T}_n$}
    \BlankLine
    
    Select the highest credibility IMU $S_I^*$\;
    \While{there are remaining sensors in the L2 sensor group}{
    Select the current highest credibility sensor $S_i^{L_2^*}$\;
        $S_i^{L_2^*} \leftarrow \arg \min C_{S_i}$\;
    Evaluate the state coverage of this sensor; if there are uncovered elements, find the highest performance sensor $S^{L_3^*}_i$ among the remaining sensors that can cover this missing state\;
        \If{$M^{L_2^*}_{s_i}(j) = 0$}{
            $S_i^{L_3^*} \leftarrow \arg \max \left(\frac{1}{E(S_i)} \cdot M_{S_i}(j)\right)$ 
            }
    If there are sensors in L3 with higher credibility than those selected in L2, then among these, select the sensor with the optimal performance index for different covered state variables and add this sensor to the selected tree structure\;
    $S^{L_3^*}=\arg \max _{S_i \in L 3}\left(\frac{1}{E(S_i)} \cdot \prod_{S_l \in L 2} 1_{\left\{C\left(S_i\right)>C\left(S^{L_2^*}\right)\right\}}\right)$ \;
    $\mathcal{T}_1$: $S_I^* \rightarrow S_i^{L_2^*} \rightarrow S^{L_3^*}$\;
    Remove the sensors included in $\mathcal{T}_1$ from the sensor group\;
    }
\end{algorithm}

\section{IMM-based Multi-Source Resilient Fusion Pipeline}\label{IMM-based Multi-Source Resilient Fusion}
This section proposes a resilient fusion algorithm based on the IMM to integrate different submodels, as the accuracy of different submodels is related to various interference factors. By utilizing the Bayesian probability process, it fully integrates the estimated state of each model to adapt to a wide range of flight conditions, achieving resilience in essence. Under outdoor open conditions with good GPS satellite reception, the GPS-based submodel is more applicable; in environments with denial or communication interference, submodels based primarily on vision and LIDAR will have superior performance. This section mainly introduces the IMM-based interactive multi-source resilient fusion framework, primarily as shown in Fig.~(\ref{ch6:fig:IMM_3GPS}). Unlike the traditional IMM algorithm, which mainly interacts between different dynamics (e.g., acceleration model, constant velocity model, etc.), and different from the ISUS model proposed by Meng et al., which includes only two types of sensors in its submodels, the submodels applied in this paper are combinations of different sensors selected through Section \ref{sensor_select}. Moreover, the internal algorithms of the submodels use a manifold-based right-invariant Kalman filter algorithm for fusion updates, which is more adaptable to the potential needs of resilient fusion in multi-source data navigation systems.
\subsection{Sensor Submodel Update Based on Right-Invariant Kalman Filter}
\begin{algorithm}[!t]
    \small
    \caption{RIEKF-Based Submodel One-Step State Update Algorithm}
    \label{RIEKF_submodel}
    \KwIn{
    The system state ${^{(m_j)}\bm{\chi}_{k-1}}$ and error state covariance ${^{(m_j)}\bm{P}_{k-1}}$ of the $j$th submodel\;
    Sensor measurements ${^{(i)}\bm{Z}}_k$ in the $j$th submodel\;
    System state transition matrix $\boldsymbol{\Phi}_{t,t+1}$, predicted state covariance matrix $\bm{Q}_0$, measurement covariance matrix ${^{(i)}\bm{R}_k}$, noise excitation matrix ${^{(i)}\bm{G}_k}$\;
    }
    \KwOut{
    The system state ${^{(m_j)}\bm{\chi}_{k}}$ and error state covariance ${^{(m_j)}\bm{P}_{k}}$ of the $j$th submodel\;
    Sensor measurement residual statistic ${^{(i)}q_k}$ in the $j$th submodel}
    \BlankLine
     Right-invariant error state one-step prediction: $d\boldsymbol{\chi}^R_{t,t+1} \longleftarrow \boldsymbol{\Phi}_{t,t+1} d\boldsymbol{\chi}^R_{t}$ \;
    One-step covariance update: $\bm{P}_{t,t+1} \longleftarrow \boldsymbol{\Phi}_{t,t+1} \bm{P}_{t} \boldsymbol{\Phi}_{t,t+1}^\top + \bm{G}_0 \bm{Q}_0 \bm{G}_0^{\top}$\;
    Compute KF gain: $\bm{K}_{t+1} \longleftarrow \bm{P}_{t,t+1} \bm{H}^{R,\top}_{t+1}\left(\bm{H}^{R}_{t+1} \bm{P}_{t,t+1} \bm{H}^{R,\top}_{t+1} \right)^{-1}$\;
    State update: $d\boldsymbol{\chi}^R_{t+1} \longleftarrow d\boldsymbol{\chi}^R_{t,t+1} + \bm{K}_{t+1} \left(\bm{z}_k - \bm{H}^{R}_{t+1}  d\boldsymbol{\chi}^R_{t,t+1} \right)$\;
    Covariance update: $\bm{P}_{t+1} \longleftarrow \left(\bm{I} -  \bm{K}_{t+1} \bm{H}^{R}_{t+1}\right)\bm{P}_{t,t+1}$\;
\end{algorithm}
The Invariant Extended Kalman Filter (IEKF) algorithm has gained widespread attention and application in recent years due to its stronger robustness to nonlinear systems and non-Gaussian noise, as well as its adaptability to complex environments \cite{Barrau2017TAC,Changwu2022,Hartley2020,ye2023semiaerodynamic}. Detailed derivations of system dynamics and measurement equations applied to the Invariant Kalman Filter are provided in previous work. The RIEKF algorithm framework for submodel updates is shown in \ref{RIEKF_submodel}.

To provide realtime feedback on the performance indexes of sensors within each submodel, we calculate the measurement innovation covariance for each sensor in the submodel. This represents the uncertainty in the difference between the measured values and the predicted measurements during the measurement update step. The right-invariant error state innovation for the $i$th sensor at time $k$, denoted as $^{(i)}\mathcal{Y}_k$, can be calculated as follows:
\begin{equation}\label{cal_innovation}
     {^{(i)}\mathcal{Y}_k} = {^{(i)}\bm{\gamma}^R_k }- {^{(i)}\bm{H}^R_k}  {^{(i)}\bm{\xi}^R_k}
\end{equation}

The corresponding innovation error covariance matrix is shown as:
\begin{equation}\label{cal_Sk}
    {^{(i)}\mathcal{S}_k} = {^{(i)}\bm{H}^R_k} {^{(i)}\bm{P}_{k/k-1}}  {^{(i)}\bm{H}^{R\top}_k} + {^{(i)}\bm{R}_k}
\end{equation}

The measurement residual statistic is calculated as:
\begin{equation}
    {^{(i)}q_k} = {^{(i)}\mathcal{Y}_k}^{\top} {^{(i)}\mathcal{S}_k}^{-1} {^{(i)}\mathcal{Y}_k}
\end{equation}

It is important to note that due to the flexibility in sensor selection within submodels, there may be differences in the choice of state variables between different submodels. This is manifested as the coexistence of public states \({^{(i)}\bm{\chi}^c}\) and private states \({^{(i)}\bm{\chi}^p}\). Public states refer to the state variables shared by all sensors, meaning these state variables appear in every submodel. Private states refer to those unique to a specific sensor, such as differences in IMU models leading to inconsistent estimates of IMU bias errors among different IMUs. In subsequent resilient interactions, it is necessary to isolate the differences between submodels, resiliently fuse the common part of the state variables, and independently update the private states of the sensors.
\subsection{Resilient Interaction}
Since each submodel is processed in parallel synchronously, we need a practical approach to fully fuse the public states \({^{(i)}\bm{\chi}^c}\) from the estimation results of these multiple submodels. The IMM (Interactive Multiple Model) algorithm, based on Bayesian theory, provides us with a theoretical foundation. For a state estimation system with \(n\) submodels, the system's state estimate \(\hat{\bm{\chi}}_{k / k}\), given the observation sequence \(\bm{Z}_k\), which is the expected value of the system state \(\bm{\chi}_k\), is obtained by weighted summation of contributions from multiple submodels.
\begin{equation}
\begin{aligned}
    \hat{\bm{\chi}}_{k / k}&=E\left[\bm{\chi}_k \mid \bm{Z}_k\right]\\
    &=\sum_{i=1}^n E\left[{^{(i)}\bm{\chi}_k} \mid {^{(i)}\bm{Z}_k}, {^{(i)} m_k}\right] P\left\{{^{(i)}m_k} \mid {^{(i)}\bm{Z}_k}\right\}
    \end{aligned}
\end{equation}
where, $m_k^{(i)}$ is shown as the $i$-th submodel.
\begin{equation}
    {^{(i)} m_k} \sim\left\{{^{(i)}\bm{R}_k}, {^{(i)}\bm{P}_{k-1 / k-1}}, {^{(i)}\bm{Q}_k}, {^{(i)}\bm{H}_k}, {^{(i)}\bm{F}_k}\right\}
\end{equation}
\({^{(i)}\bm{Z}_k}\) represents the set of measurements for the model, which can be expressed as:
\begin{equation}
    \bm{Z}_k=\left[\begin{array}{lllll}
   { ^{(1)}\bm{Z}_k} & \cdots & {^{(i)}\bm{Z}_k} & \cdots & {^{(n)}\bm{Z}_k}
    \end{array}\right]^{\top}
\end{equation}
\(P\left\{{^{(i)}m_k} \mid {^{(i)}\bm{Z}_k}\right\}\) is the posterior probability of the model. To update the posterior mode probability, the likelihood function of the multivariate Gaussian distribution given the conditions is used to describe:
\begin{equation}
\begin{aligned}
    {^{(i)}L_k} & \triangleq P\left[\tilde{{^{(i)}\bm{Z}}_k} \mid {^{(i)}m_k}, {^{(i)}\bm{Z}_{k-1}}\right] \\
    & =\frac{1}{\sqrt{\operatorname{det}\left(2 \pi {^{(i)}\bm{S}_k}\right)}} \exp \left[-\frac{1}{2} {^{(i)}\tilde{\bm{Z}} }_k^{\top} {^{(i)}\bm{S}_k^{-1}} {^{(i)}\tilde{\bm{Z}}_k}\right]
\end{aligned}
\end{equation}
The likelihood value \({^{(i)}L_k}\) represents the probability of observing the current measurement value given the model and its prediction. A high likelihood value indicates that the model prediction is closer to the actual observation. Therefore, a high likelihood value for a submodel is considered to indicate that the submodel is more reliable and has higher credibility.

For an individual submodel, which internally alternates updates among several sensors, constructing a joint observation model is necessary when calculating the likelihood of the entire submodel. The joint observation model matrix for the \(i\)th submodel is as follows:
\begin{equation}
    {^{(i)}\bm{H}_k}=\left[\begin{array}{lllll}
   { ^{(i)}\bm{H}_1} & {^{(i)}\bm{H}_2} & \cdots & {^{(i)}\bm{H}_n}
    \end{array}\right]^{\top}
\end{equation}
by substituting \({^{(i)}\bm{Z}_k}\) and \({^{(i)}\bm{H}_k}\) into Eq.~(\ref{cal_innovation}) and Eq.~(\ref{cal_Sk}), the joint innovation and the innovation covariance matrix can be calculated. In Eq.~(\ref{cal_Sk}), \({^{(i)}\bm{R}_k}\) should be replaced with the joint measurement covariance matrix \({^{(i)}\bm{R}_{comb}} = \operatorname{diag} \,\left[
\begin{array}{ccc}
   {^{(i)}\bm{R}_1} & \cdots & {^{(i)}\bm{R}_n} \\
\end{array}\right]\).

Therefore, according to Bayesian theory, to achieve normalization, the predicted prior probability for the \(i\)th submodel can be represented as follows:
\begin{equation}
    {^{(i)}\mu_k}=\frac{{^{(i)}\mu_{k / k-1}} {^{(i)}L_k}}{\sum_{j=1}^M {^{(j)}\mu_{k / k-1}} {^{(j)}L_k}}
\end{equation}
The previous step state \(\bm{\chi}_{k}\) of each submodel is combined with the interactive weights in the IMM filter. The interactive weights for the submodels can be written as:
\begin{equation}
{^{(j \mid i)}\mu_{k-1 \mid k-1}}=\frac{1}{^{(i)}\hat{\mu}_{k \mid k-1}} \pi_{j i} {^{(j)}\mu_{k-1}}, \quad i, j=1,2,...n
\end{equation}
where, \(\mu_{k-1}^j\) is the probability of the \(j\)th submodel at the previous step \(k-1\), and \(\pi_{ji}\) is the model transition probability from the \(j\)th submodel to the \(i\)th submodel. The sum of the interactive probabilities for the submodels is:
\begin{equation}
    {^{(i)}\hat{\mu}_{k \mid k-1}}=\sum_{j=1}^n \pi_{j i} {^{(j)}\mu_{k-1}} \quad i=1,2
\end{equation}

The state transition matrix of the IMM filter is controlled by a first-order Markov assumption, meaning the current state contains all the information needed to characterize the probability distribution of the next step. The model transition probabilities \(\pi_{ji}\) are combined into a transition matrix and are assumed to be known a priori. Each element of the transition matrix \(\pi_{ji}\) represents the probability of transitioning from model \(i\) to model \(j\). The transition matrix \(\pi_{ji}\) is as follows:
\begin{equation}
    \pi_{ji} = \left[\begin{array}{cccc}
        \pi_{1,1} & \pi_{1,1} & \cdots & \pi_{1,i} \\
        \pi_{2,1} & \pi_{2,1} & \cdots & \pi_{2,i} \\
        \vdots & \vdots&\ddots&\vdots\\
        \pi_{j,1} & \pi_{j,1} & \cdots & \pi_{j,i} \\
    \end{array}\right]
\end{equation}

The interactive state and covariance matrix updates for each submodel can be represented as follows:

\begin{equation}
\resizebox{1\linewidth}{!}{$
\begin{aligned}
{^{(i)}\hat{\bm{\chi}}_{k / k}} & =\sum_{j=1}^M {^{(i)} \hat{\bm{\chi}}_{k / k}}{^{(j \mid i)}\mu_k} \\
{^{(i)}\hat{\bm{P}}_{k / k}} & =\sum_{i=1}^M\left[{^{(j)}\hat{\bm{P}}_{k / k}}+\left(^{(i)}{\bm{X}}_{k / k}-{^{(j)}\hat{\bm{X}}_{k / k}}\right)\left({^{(i)}\hat{\bm{\chi}}_{k / k}}-{^{(j)}\hat{\bm{\chi}}_{k / k}}\right)^T\right] {^{(j\mid i)}\mu_k}
\end{aligned}$}
\end{equation}

Finally, after the measurement updates of each submodel, the overall system state and covariance optimal estimates can be obtained by fusing the interactive weights of all submodels:
\begin{equation}
\resizebox{1\linewidth}{!}{$
\begin{aligned}
    & \hat{\bm{\chi}}_{k / k}=\sum_{i=1}^M {^{(i)}\hat{\bm{\chi}}_{k / k}} \mu_k^{(i)} \\
    & \hat{\bm{P}}_{k / k}=\sum_{i=1}^M\left[{^{(i)}\hat{\bm{P}}_{k / k}}+\left(\hat{\bm{\chi}}_{k / k}-{^{(i)}\hat{\bm{\chi}}_{k / k}}\right)\left(\hat{\bm{\chi}}_{k / k}-{^{(i)}\hat{\bm{\chi}}_{k / k}}\right)^{\mathrm{T}}\right] {^{(i)}\mu_k}
\end{aligned}$}
\end{equation}

\subsection{Model Transition Probability}
In the IMM-based resilient interaction algorithm, the model transition probability matrix \(\pi\) serves as a critical component. This matrix provides the ideal probabilities for switching from one submodel to another, based on the prior information about the system dynamics, reflecting the possibility of dynamic transitions between models while pre-adjusting the weights of each model. The model transition probabilities \(\pi\) together with the model likelihood functions \({^{(i)}L_k}\) determine the weights for actual model transitions at each step. The likelihood function assesses the adaptability of each model to the current observational data, where a high likelihood value implies that the model can better explain the current observations, and the model weights are updated at each timestep based on new observation innovations and innovation covariance.

Therefore, determining the model transition probabilities \(\pi\) is a key task. Inappropriate \(\pi\) values can compromise the system's accuracy and affect the state tracking response of the system. The goal of this section is to propose an algorithm for dynamically adjusting the allocation of the state transition matrix at each moment based on the credibility performance indexes of the sensors.

Section \ref{sensor_reliability_index} defines the credibility index \(^{(i)}H_n\) for a group of sensors, which is composed of expected and actual performance indexes. When a sensor's accuracy is higher, its credibility is higher, and the value is larger. Therefore, submodels corresponding to sensors with higher accuracy should be more credible, and thus the probability of maintaining their current state should be higher. However, the multi-source resilient fusion interactive algorithm is built on the interaction between submodels, necessitating a unified description of heterogeneous sensors within submodels to calculate submodel credibility weights. This unified description involves not only converting outputs from different types and functions of sensors into a comparable format but also includes a comprehensive evaluation of each sensor's performance to ensure the accuracy and fairness of weight distribution. The calculation of model transition probabilities mainly involves the following steps:

(1) Calculate submodel weights: Define that the navigation system is divided into \(n\) submodels, each containing \(m\) sensors. The credibility index of the \(j\)th sensor in the \(i\)th submodel is denoted as \(^{(i)}H_{j}\). The average method is used to calculate the weight of each submodel:
\begin{equation}
    {^{(i)}\omega} = \frac{^{(i)}H_{1} + ^{(i)}H_{2}+...+^{(i)}H_{j}}{m}
\end{equation}

After obtaining the weights for all submodels, normalize them:
\begin{equation}
    {^{(i)}\overline{\omega}} = \frac{{^{(i)}\omega}}{\sum^{n}_{i = 1} {^{(i)}\omega}}
\end{equation}

(2) Calculate the self-transition probability \(\pi_{ii}\)
\begin{equation}
    \pi_{ii} = b \cdot {^{(i)}\overline{\omega}} 
\end{equation}
where, \(b\) is defined as the baseline probability, reflecting the tendency of each model to maintain its current state when not influenced by external observations.

(3) Calculate the transition probabilities \(\pi_{ij}\) between submodels, where \(j \neq i\).

The remaining probability is evenly distributed among the other \(n-1\) models, i.e., \(\pi_{ij} = \frac{1-\pi_{ii}}{n-1}\). This even distribution helps provide a fair "competition" environment for the remaining submodels, especially when it is difficult to accurately evaluate the transition probability of each model.
\begin{equation}\label{ch6:eq:cal_pi}   
\pi =
\begin{bmatrix}
b \cdot {^{(1)}\overline{\omega}}  & \frac{1 - b \cdot {^{(1)}\overline{\omega}} }{n-1} & \cdots & \frac{1 - b \cdot {^{(1)}\overline{\omega}} }{n-1} \\
\frac{1 - b \cdot {^{(2)}\overline{\omega}} }{n-1} & b \cdot {^{(2)}\overline{\omega}}  & \cdots & \frac{1 - b \cdot {^{(2)}\overline{\omega}} }{n-1} \\
\vdots & \vdots & \ddots & \vdots \\
\frac{1 - b \cdot {^{(n)}\overline{\omega}} }{n-1} & \frac{1 - b \cdot {^{(n)}\overline{\omega}} }{n-1} & \cdots & b \cdot {^{(n)}\overline{\omega}} 
\end{bmatrix}
\end{equation}
\section{Redundant IMU/GPS Flight Data Validation}\label{exp_on_multi_GPS/INS}
This section validates the proposed resilient algorithm using a conventional layout drone equipped with redundant sensor configurations. The drone is outfitted with multiple navigation systems, including a high-precision Fiber Optic Gyro Inertial Measurement Unit (FOG IMU), two different models of RTK GPS, and a standard single-point Global Positioning System (GPS). The combination of these devices ensures high precision and reliable positioning and navigation under various flight conditions, serving to verify the multi-source navigation resilient fusion algorithm. The drone is turbojet-powered, with an empty fuel weight of about 110 kg and a cruising speed of about 50 m/s. The accuracy of the sensors is as shown in Table \ref{ch6:tab:sensor}.
\begin{table}[tbp]
    \centering
    \small
    \begin{threeparttable}[b]
    \caption{Sensors Specifications for the UAV with Redundancy\tnote{a}}
    \label{ch6:tab:sensor}
    \begin{tabular}{cccc}
    \toprule
       \textbf{Sensor}  & \textbf{Parameter} & \textbf{Specification} & \textbf{Unit}\\
    \midrule
       IMU  & Bias Stability & $0.1$ & deg/hr\\
       \midrule
       \multirow{2}{*}{RTK 1}   & Positioning Accuracy & $0.02$ & m\\
       & Speed Accuracy & $0.03$ & m/s\\ 
       \midrule
       \multirow{2}{*}{RTK 2}   & Positioning Accuracy &  $0.03$ & m\\
       & Speed Accuracy & $0.02$ & m/s\\ 
       \midrule
       \multirow{2}{*}{GPS}    & Positioning Accuracy &  $0.5$ & m\\
       & Speed Accuracy & $0.1$ & m/s\\ 
       \bottomrule
    \end{tabular}
    \begin{tablenotes}
    \item[a] Due to the sensitivity of the data, specific sensor models cannot be provided.
    \end{tablenotes}
    \end{threeparttable}
\end{table}

\begin{figure}[!t]
    \centering
    \includegraphics[width=1\linewidth]{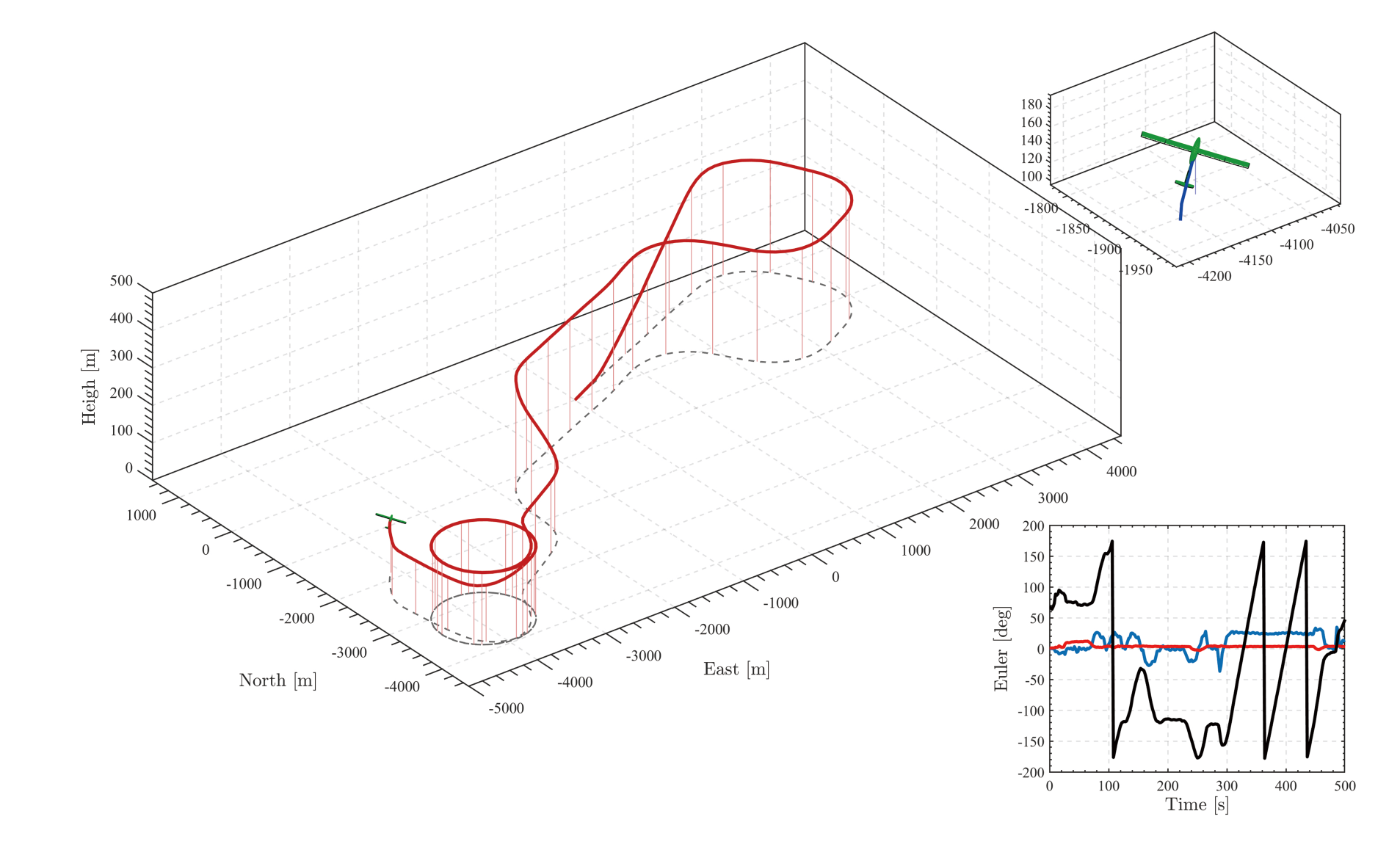}
    \caption{UAV flight experiment 3D trajectory and attitude display.}
    \label{ch6:fig:Traj}
\end{figure}

To further evaluate the performance of the algorithm, this section employs IMU + 3GPS for the aforementioned resilient fusion for algorithm validation and uses IMU + RTK GPS 1 (where RTK GPS 1 has the highest accuracy) for a comparative experiment. The comparative experiment adopts the advanced Error State Right-Invariant Extended Kalman Filter (ES-RIEKF) algorithm, which has been extensively theoretically and experimentally validated for its superior performance. We use the pose results from the flight control combination navigation output as a reference benchmark to compare the advantages of different algorithms. The flight trajectory of the flight experiment is shown in Fig.~(\ref{ch6:fig:Traj}). This study designs two sets of experiments to verify the "resilience" capability of the proposed algorithm. The first experiment is conducted under normal sensor operation conditions, aiming to verify the algorithm's accuracy and robustness under normal conditions. The second set of experiments introduces heavy-tailed noise errors into the GPS measurement data and sets up a jamming environment to simulate scenarios where sensor performance is impaired, aiming to examine the adaptability and robustness of the algorithm when facing fluctuations in data output quality.

Firstly, a description of the design of the filter model is given as following.
\subsection{System Resilience Filtering Model Design}
The algorithm design employs a 15-dimensional full system nominal state variable defined as follows:
\begin{equation}
    \bm{\chi}=\left( \bm{R}_b^n,\bm{v}^n,\bm{p}^n,\bm{b}^n_g,\bm{b}^n_a \right)
\end{equation}
where \(\left(\bm{R}_b^n,\bm{v}^n,\bm{p}^n\right) \in \mathbb{SE}_2(3)\) represents the pose and velocity of the vehicle, and \(\left(\bm{b}^n_g,\bm{b}^n_a\right) \in (\mathbb{R}^3)^2\) are the three-axis zero-bias errors of the IMU's gyroscope and accelerometer.

\begin{figure*}[tbp]
    \centering
    \includegraphics[trim={0.1cm 0.1cm 0.1cm 0.1cm},clip,width=0.8\linewidth]{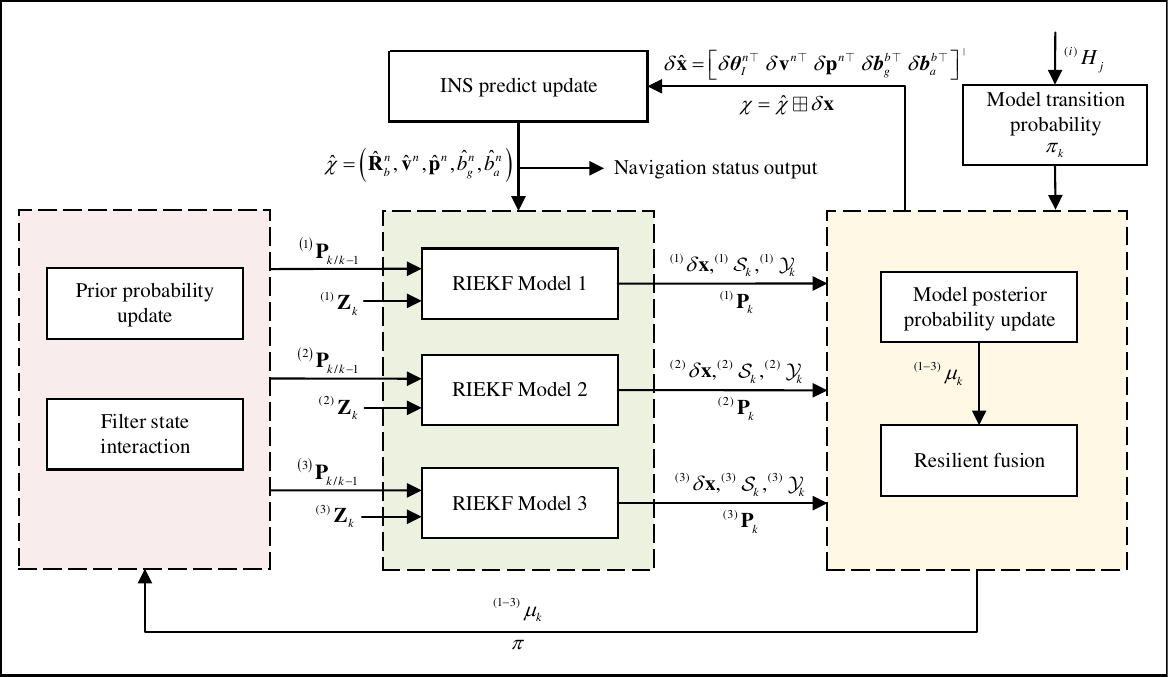}
    \caption{IMU/GPS fusion framework based on IMM interactive redundancy.}
    \label{ch6:fig:IMM_3GPS}
\end{figure*}

The system framework for the IMM-based interactive redundancy IMU/GPS fusion, as shown in \ref{ch6:fig:IMM_3GPS}, is divided into four main parts: (1) Submodel prior interaction, (2) Submodel measurement update, (3) Model posterior update and resilient fusion, (4) INS prediction update and state feedback correction.

Submodels and the prediction model adopt the advanced Error-State Right-Invariant Extended Kalman Filter (ES-RIEKF) algorithm based on manifolds, aimed at enhancing convergence capability and estimation accuracy under disturbances. The system state is represented by the error state \(\delta \bm{x}\),
\begin{equation}
\delta \bm{x}=  \left[\begin{array}{ccccc}
\delta \boldsymbol{\theta}_I^{n \top} & 
\delta \bm{v}^{n \top} & 
\delta \bm{p}^{n \top}   &
\delta \bm{b}_g^{b \top} & 
\delta \bm{b}_a^{b \top} 
\end{array}\right]^{\top}
\end{equation}

The sensors are divided into three submodels, which interact through a three-submodel resilient fusion process. The estimated error state \(\delta \hat{\bm{x}}\) is fed back into the INS prediction system for error correction:
\begin{equation}
    \bm{\chi}=\hat{\bm{\chi}} \boxplus \delta \bm{x} = \hat{\bm{\chi}} \exp{(\delta \bm{x})}
\end{equation}
where, \(\boxplus\) represents the addition operator defined on the manifold.

The measurements for the three submodels in the system correspond to the velocity and position measured by three different GPS devices. The measurement vectors are as follows:
\begin{equation}
    {^{(i)}\bm{Z}_k} = \left[ \begin{array}{cc}
       {^{(i)}\bm{p}^n}^{\top}  &  {^{(i)}\bm{v}^n_i}^{\top}\\
    \end{array} \right]^{\top}
\end{equation}

\begin{table}[!t]
    \caption{Filter parameters setting.}
    \centering
    \small
    \begin{tabular}{clcc}
    \toprule
       \textbf{Variable}& \textbf{Parameter}&\textbf{Value}  & \textbf{Unit}\\
       \midrule
       \multirow{4}{*}{$\bm{Q}_0$} 
       & $\sigma_a^2$ & $(10^{-3})^2$ & $\left(\mathrm{m}/\mathrm{s}^2\right)^2$\\
       & $\sigma_g^2$ & $(10^{-4})^2$ & $\left(\mathrm{rad}/\mathrm{s}\right)^2$\\
       & $\sigma_{\bm{b}_a}^2$ & $(10^{-4})^2$ & $\left(\mathrm{m}/\mathrm{s}^2\right)^2/\mathrm{Hz}$ \\
       & $\sigma_{\bm{b}_g}^2$ & $(2\times10^{-5})^2$ & $\left(\mathrm{rad}/\mathrm{s}\right)^2/\mathrm{Hz}$ \\
       \midrule
       \multirow{5}{*}{$\bm{P}_0$}
       & $\sigma_{\delta \boldsymbol{\theta}_I}^2$ & $([1;1;5]/57.3)^2$ & $\mathrm{rad}^2$\\
       & $\sigma_{\delta \bm{v}}^2$ & $(0.1)^2$ & $(\mathrm{m}/\mathrm{s})^2$\\
       & $\sigma_{\delta \bm{p}}^2$ & $(0.2)^2$ & $\mathrm{m}^2$\\
       & $\sigma_{\delta \bm{b}_g}^2$ & $(4.8478e-5)^2$ & $(\mathrm{rad}/\mathrm{s})^2$\\
       & $\sigma_{\delta \bm{b}_a}^2$ & $(0.05)^2$ & $(\mathrm{m}/\mathrm{s}^2)^2$\\
       \midrule
       \multirow{2}{*}{$^{(1)}\bm{R}_0$}
       & $\sigma_{\bm{v}_G}^2$ & $(0.01)^2$  & $(\mathrm{m}/\mathrm{s})^2$\\
       & $\sigma_{\bm{p}_G}^2$ & $(0.01)^2$ & $(\mathrm{m})^2$\\
       \midrule
       \multirow{2}{*}{$^{(2)}\bm{R}_0$}
       & $\sigma_{\bm{v}_G}^2$ & $(0.02)^2$  & $(\mathrm{m}/\mathrm{s})^2$\\
       & $\sigma_{\bm{p}_G}^2$ & $(0.02)^2$ & $(\mathrm{m})^2$\\
       \midrule
       \multirow{2}{*}{$^{(3)}\bm{R}_0$}
       & $\sigma_{\bm{v}_G}^2$ & $(0.02)^2$  & $(\mathrm{m}/\mathrm{s})^2$\\
       & $\sigma_{\bm{p}_G}^2$ & $(0.02)^2$ & $(\mathrm{m})^2$\\
       \bottomrule
    \end{tabular}
    \label{ch6:tab:Filter parameters}
\end{table}

\subsection{Sensor System Performance Evaluation}
In actual flight experiments, the UAV flies in open areas with all onboard GPS devices functioning normally. To evaluate the system performance and algorithmic correctness of the redundancy system when sensors are operating as expected, offline data simulation of 1IMU+3GPS flight data is conducted. This is done to assess the accuracy and robustness of the navigation system. The core parameters of the system filter are set as shown in Section \ref{ch6:tab:Filter parameters}.
\subsubsection{Model transition probabilities calculations}
Based on the sensor accuracy specifications from Table \ref{ch6:tab:sensor} and using Eq.~(\ref{ch6:eq:sensor_err_cal}), the expected performance indexes for the three GPS units can be calculated, where \(t=0.1s\) is taken from the GPS update cycle: \(^{(1)}E = 1/0.023=43.478\) (m), \(^{(2)}E = 1/0.032 = 31.25\) (m), \(^{(3)}E = 1/0.505 = 1.9802\) (m). During the flight, the RTK/GPS functions normally with DOP values all less than 1, \(^{(1,2,3)}P = 10\), then the system credibility index is:
\begin{equation}
    ^{(1)}H = 434.78 (\text{m}),^{(2)}H = 312.5 (\text{m}),^{(3)}H = 19.802 (\text{m})
\end{equation}

Given the baseline probability \(b = 1\), according to Eq.~(\ref{ch6:eq:cal_pi}), the calculation yields:
\begin{equation}
\pi = 
\begin{bmatrix}
    0.5668 & 0.2166 & 0.2166 \\
    0.2963 & 0.4074 & 0.2963 \\
    0.4871 & 0.4871 & 0.0258 \\
\end{bmatrix}
\end{equation}

\subsubsection{Results}
\begin{figure}[!t]
    \centering
        \includegraphics[width=0.95\linewidth]{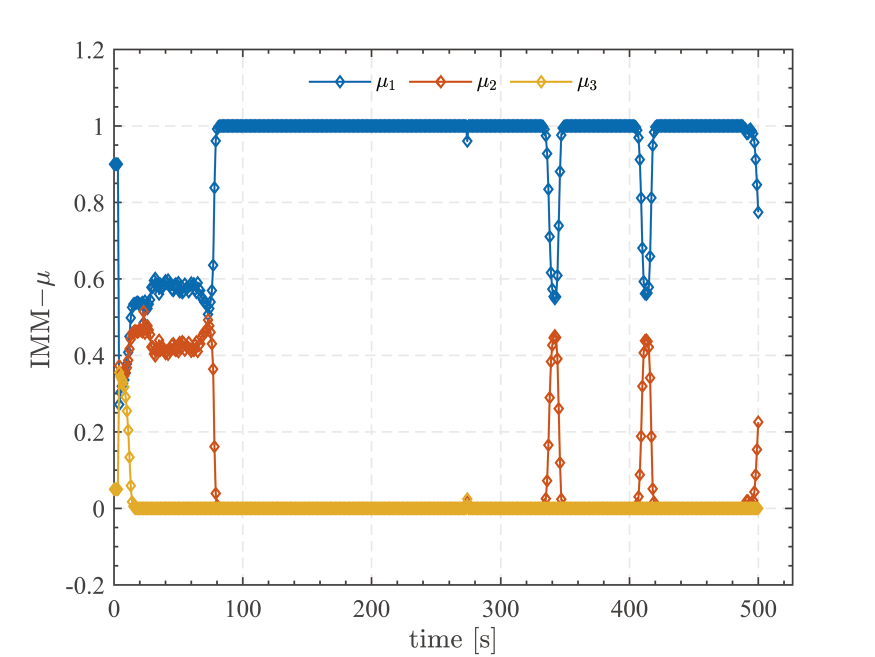}
        \caption{IMM posterior probability result.}
        \label{ch6:fig:mu_3GPS}
        \centering
        \includegraphics[width=0.95\linewidth]{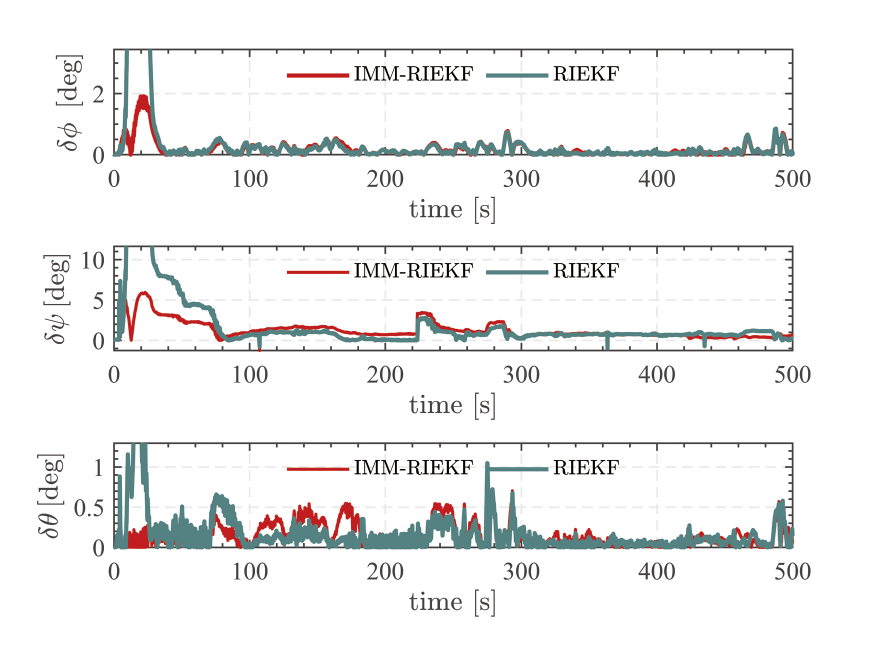}
        \caption{IMM attitude error comparison.}
        \label{ch6:fig:atterr_3GPS}
\end{figure}

\begin{table*}[!t]
    \small
    \centering
    \caption{Comparison of precision indexes of pose estimation.}
    \begin{tabular}{ccccccccc}
    \toprule
    \multirow{2}{*}{\textbf{Axial}} & \multicolumn{2}{c}{\textbf{Position STD(m)}} & \multicolumn{2}{c}{\textbf{Rotation STD(deg)}} & \multicolumn{2}{c}{\textbf{Position RMSE(m)}} & \multicolumn{2}{c}{\textbf{Rotation RMSE(deg)}}\\
    \cmidrule { 2 - 9 }
    & ES-RIEKF & PROP & ES-RIEKF  & PROP& ES-RIEKF & PROP & ES-RIEKF  & PROP\\
    \midrule
    North (N)/Roll &  0.094 & 0.099 &  0.358&  0.185&0.171&0.174& 1.157&0.323\\
    East (E)/Pitch &  0.117& 0.119 &0.160& 0.149 &  0.202 & 0.202 & 0.301 &0.207 \\
    Down (D)/Yaw & 0.015 &0.024 &2.053 & 1.273 & 0.022 &   0.039 &4.829 &1.662\\
    \bottomrule
    \end{tabular}
    \label{ch6:tab:rotpos_err_rmse}
\end{table*}
Fig.~(\ref{ch6:fig:mu_3GPS}) shows the posterior probability curves of the three submodels during the interactive fusion period. It can be observed that the posterior probability of Submodel 1 remains at a higher level for most of the time and nearly reaches 100\% after 80 seconds. This phenomenon indicates that Submodel 1 performs excellently in terms of accuracy and credibility, aligning with expectations. Meanwhile, the posterior probability of Submodel 2 largely complements that of Submodel 1, whereas Submodel 3's probability quickly drops to near zero after 30 seconds, suggesting it is eliminated by the fusion algorithm. This outcome demonstrates the effectiveness of the fusion algorithm in evaluating and comparing the performance of different submodels, accurately identifying and relying on the most reliable data sources to optimize the overall system performance.

\begin{figure}[!t]
    \centering
    \includegraphics[width=1\linewidth]{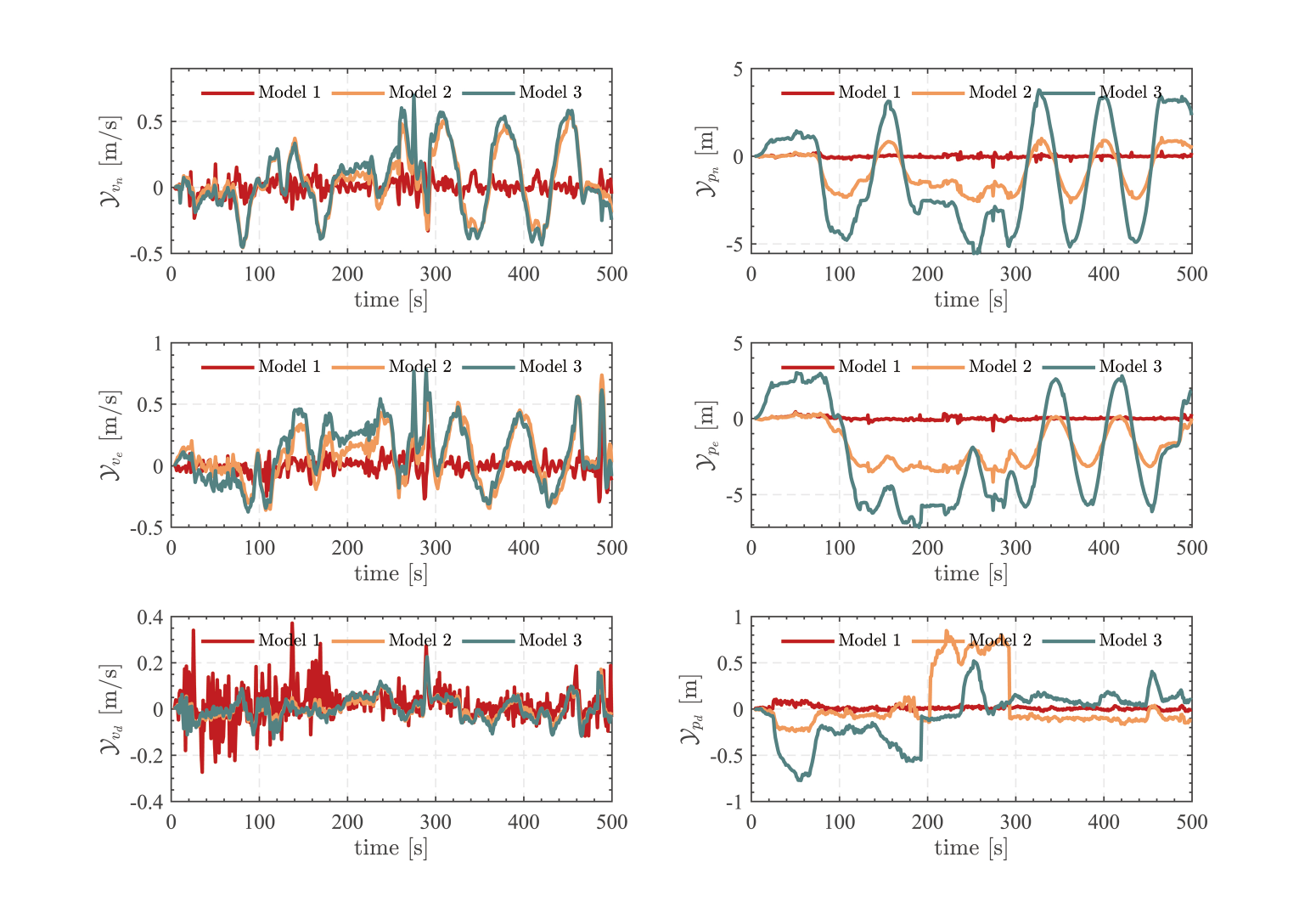}
    \caption{The comparison of each submodel innovation.}
    \label{ch6:fig:innovation_3GPS}
\end{figure}

Similar conclusions are drawn in Fig.~(\ref{ch6:fig:innovation_3GPS}), which showcases the innovation of measurements for submodels. The three graphs on the left compare the three-axis velocity components of the measurement innovation, and the three on the right compare the three-axis position components. It is evident that Submodel 1 has the smallest measurement innovation in all aspects except the vertical velocity. This is because there is a need to balance the accuracy of Submodel 1 in vertical positioning, leading to a slight compromise in vertical speed for Submodel 1, which is also a result of resilience.

Fig.~(\ref{ch6:fig:atterr_3GPS}) shows the comparison between IMM-RIEKF and single-model RIEKF combination results. It is evident that the advantage of IMM-RIEKF lies in its robustness, showing superior performance during the initial attitude angle convergence period. Table \ref{ch6:tab:rotpos_err_rmse} presents a comparison of pose STD and RMSE accuracy. The formulas for MAE (Mean Absolute Error) and RMSE (Root Mean Square Error) are as follows:
\begin{equation}
\begin{gathered}
    \operatorname{MAE}(k)=\frac{1}{N} \sum_{i=1}^N\left|\hat{x}_k-x_k\right|\\
    \operatorname{RMSE}(k)=\sqrt{\frac{1}{N} \sum_{i=1}^N\left(\hat{x}_k-x_k\right)^2}
\end{gathered}
\end{equation}
From Table \ref{ch6:tab:rotpos_err_rmse}, it is evident that the IMM-ESEKF algorithm proposed in this paper has a significant advantage in terms of three-axis attitude RMSE. The error metrics on the three-axis position level are very close, within the centimeter range. This experiment demonstrates that the IMM-ESEKF algorithm, by fusing three different models of GPS with varying accuracies, can effectively enhance estimation accuracy and robustness when all sensors are functioning normally.
\subsection{Robustness Evaluation Under Sensor Faults Injection}
\subsubsection{Model Setting}
This section primarily evaluates the accuracy of the Interactive Multi-Source Resilient Fusion Algorithm under conditions where the accuracy of some sensors has deteriorated to the point of failure. Given that GPS is prone to heavy-tailed non-Gaussian noise in complex environments due to multipath effects, ionospheric and tropospheric delays, among other reasons, and can also be obstructed from obtaining precise positioning information in certain frequency bands, this study employs artificially injected errors to verify the "resilience" of the algorithm. A detailed fault injection scheme has been designed, as shown in Table \ref{ch6:tab:sensor_fault_inject_set}:
\begin{table*}[!t]
    \centering
    \caption{Multi-Sensor Fault Injection Settings.}
    \begin{threeparttable}[b]
        \setlength{\tabcolsep}{7mm}{
        \begin{tabular}{cccc}
        \toprule
        \textbf{Time} & \textbf{Sensor} & \textbf{Fault Type} & \textbf{Parameters}\tnote{a} \\
        \midrule
        $100s-160s$ & GPS 1 & Heavy-tailed Noise & $p = 0.1, s = [ \begin{array}{ccc}
            2.5 & 2.5 & 0.5 \\
        \end{array}]^{\top}$\\
        $160s-260s$ &  GPS 1  & Heavy-tailed Noise & $p = 0.05, s = [ \begin{array}{ccc}
            5 & 5 & 1 \\
        \end{array}]^{\top}$\\
        $300s-500s$ & GPS 1 & Jamming & GPS 1 fixtype = 0\\
        $200s-300s$ & GPS 2 & Jamming & GPS 2 fixtype = 0\\
        \bottomrule
    \end{tabular}}
    \begin{tablenotes}
    \item[a] $p$ is the occurrence probability of heavy-tailed noise, $s$ is the scaling factor for the standard deviation of the three-axis heavy-tailed noise.
    \end{tablenotes}
    \end{threeparttable}
    \label{ch6:tab:sensor_fault_inject_set}
\end{table*}

The IMM-RIEKF algorithm proposed in this paper is compared with single-model algorithms that have the same filter parameter settings: RIEKF Model 1\footnote{Model 1 indicates the combination of IMU and GPS 1 sensors, abbreviated as RIEKF M1, and similarly for subsequent models}, RIEKF Model 2, and ESEKF Model 1. The model settings are shown in Table \ref{ch6:tab:IMM_Model_set}:
\begin{table*}[!t]
    \centering
    \caption{Configuration of the Control Model for Sensor Error Injection Experiment.}
    \begin{tabular}{p{2.5cm}p{6cm}p{6cm}}
    \toprule
      \textbf{Model}   &  \textbf{Configuration} & \textbf{Purpose}\\
      \midrule
       IMM-RIEKF  & No additional settings &  \\
       \midrule
       RIEKF M1  & 300-500s GPS 1 2. Switch position signal directly to GPS after jamming. & Compare accuracy and robustness under error injection and switching.\\
        \midrule
       RIEKF M2  & 200-300s GPS 2 jamming period model without GPS correction. & Compare pose estimation performance under jamming.\\
       \midrule
       ESEKF M1  & 300-500s GPS 1 2. Switch position signal directly to GPS 2 after jamming. & Compare ESEKF and RIEKF pose accuracy and disturbance resistance.\\
       \bottomrule
    \end{tabular}
    \label{ch6:tab:IMM_Model_set}
\end{table*}
\subsubsection{Results}
\begin{figure}[!t]
    \centering
    \includegraphics[trim={0.1cm 0.1cm 0.1cm 0.1cm},clip,width=0.95\linewidth]{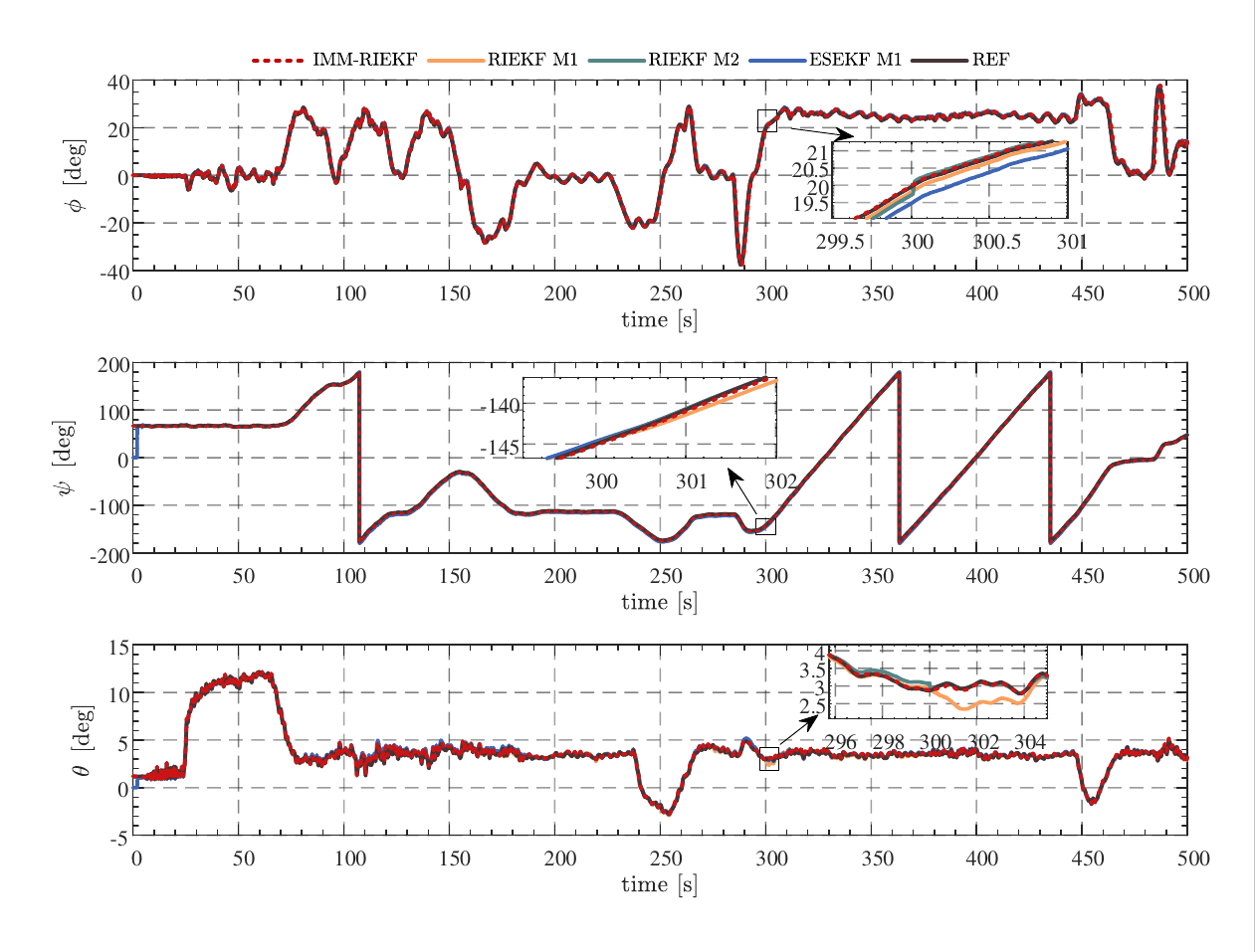}
    \caption{Comparison of triaxial attitude solution results for the model after sensor error injection.}
    \label{ch6:fig:addnoise_att}
    \includegraphics[trim={0.1cm 0.1cm 0.1cm 0.1cm},clip,width=0.95\linewidth]{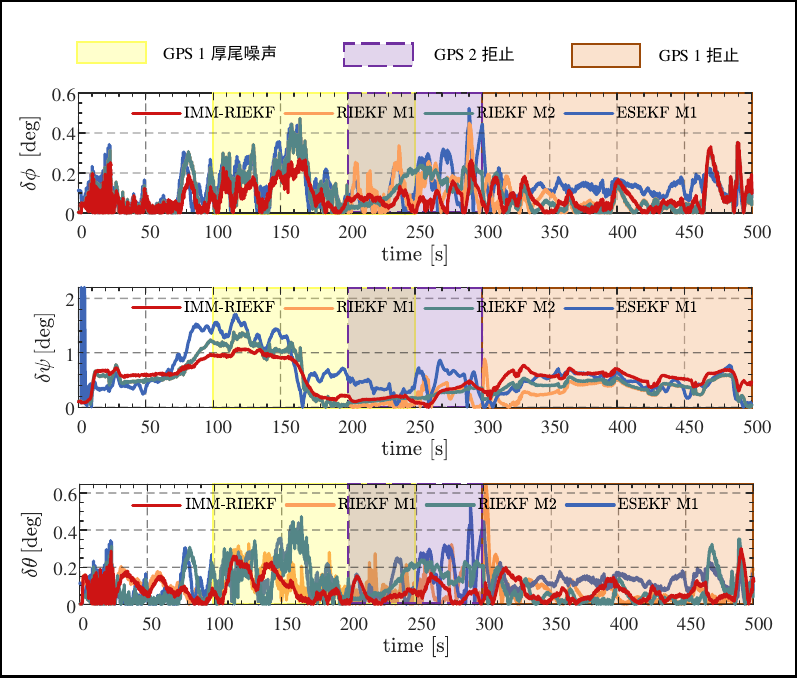}
    \caption{Comparison of triaxial attitude solution errors in the model after sensor error injection.}
    \label{ch6:fig:addnoise_atterr}
\end{figure}
\begin{figure}[!t]
    \centering
        \centering
        \includegraphics[trim={0.1cm 0.1cm 0.1cm 0.1cm},clip,width=\linewidth]{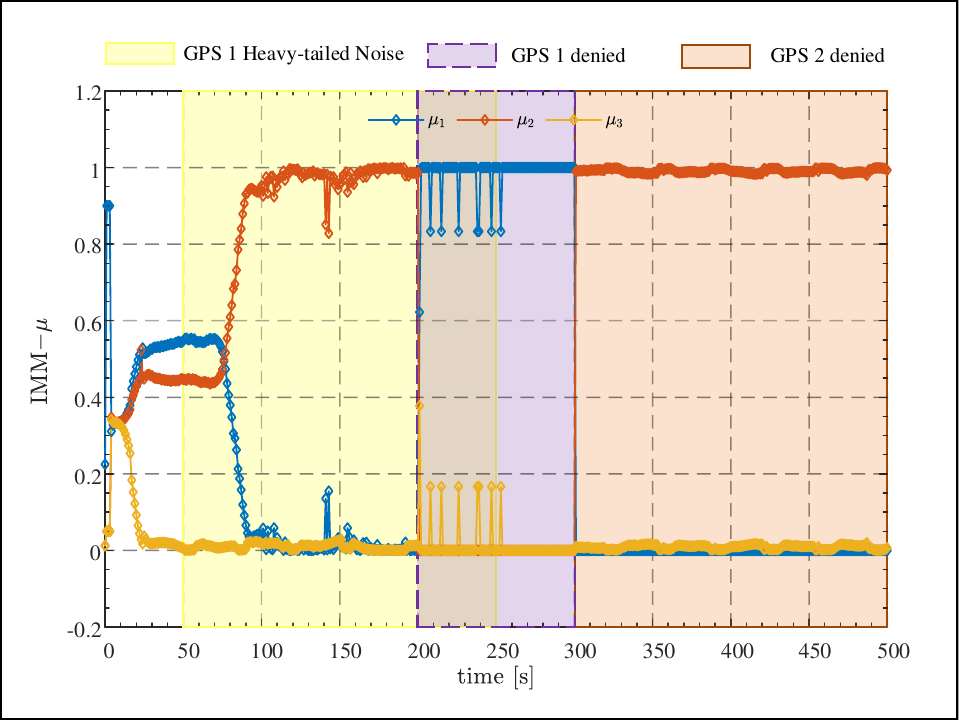}
        \caption{Switching results of IMM posterior probability ${^{(i)}\mu_k}$ after sensor error injection.}
        \label{ch6:fig:add_noise_mu_3GPS}
\end{figure}

\begin{figure}[!t]
    \centering
    \includegraphics[trim={0.1cm 0.1cm 0.1cm 0.1cm},clip,width=1\linewidth]{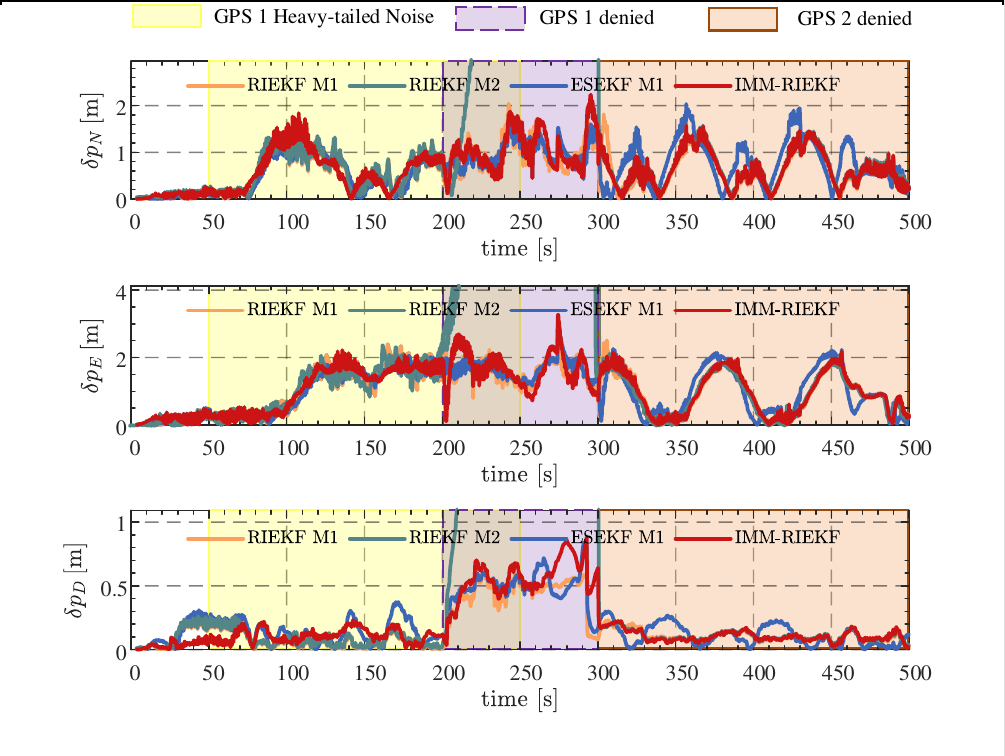}
    \caption{Comparison of triaxial attitude results in the model after sensor error injection.}
    \label{ch6:fig:addnoise_poserr}
\end{figure}
Fig.~(\ref{ch6:fig:addnoise_att}) and Fig.~(\ref{ch6:fig:addnoise_atterr}) show the comparison of three-axis attitude solutions and the comparison of errors in three-axis attitude solutions, respectively. Fig.~(\ref{ch6:fig:add_noise_mu_3GPS}) presents the IMM posterior probability \({^{(i)}\mu_k}\) switching results after sensor error injection. Fig.~(\ref{ch6:fig:addnoise_heavynoise}) shows the comparison curves between the non-Gaussian heavy-tailed white noise applied at the altitude level and the model-calculated altitude. It is observable that during the phase of heavy-tailed noise application to GPS 1, the three-axis attitude error performs better than using the RIEKF coupling model alone. In the face of heavy-tailed noise, relying on the filter's own anti-disturbance capability and the resilience advantage of IMM, Submodel 2 takes a dominant position. However, compared to the RIEKF model with integrated GPS 2 (represented by the green curve, where GPS 2 was not subjected to non-Gaussian noise during the current period), it achieves higher accuracy.

When GPS 2 denied, the weight of Submodel 1 rapidly increases. With the influence of heavy-tailed noise, Submodel 3 effectively compensates for the errors of Submodel 1 at the peak noise points, especially in the interval from 200s to 300s. At 300s, when GPS 1 experiences jamming, the RIEKF M1 single model involves switching from GPS 1 data to GPS 2 data. It is evident that at the moment of direct measurement switching, there is a jump in attitude, especially in pitch and yaw angles, with a jump of about 0.6 degrees. However, the IMM RIEKF algorithm utilizes its resilience advantage to achieve seamless switching, with no significant error jumps observed.

We also notice that under the same filter parameters, the error fluctuations of the ESEKF M1 model (represented by the blue curve) are greater than those of the RIEKF based on the invariant Kalman filter. This reflects the superior robustness of the Invariant Error EKF compared to the ESEKF. Fig.~(\ref{ch6:fig:addnoise_poserr}) shows the comparison of three-axis position errors. The differences in position errors between different models are relatively small, showing strong similarity. This is because position measurements have stronger observability. With the high accuracy of GPS, the position estimation errors for all models converge to a high standard.

This experiment demonstrates that the IMM-based redundant interactive resilient fusion framework proposed in this chapter, without employing any sensor fault diagnosis and isolation logic, without setting any complex detection timing or thresholds, solely leverages the adaptive resilience advantage between models to achieve interactive seamless switching between submodels. This enhances the adaptability of the state estimation system in complex environments, maximally utilizing the advantages of sensors to achieve a complementary enhancement of pose accuracy and robustness.
\section{Discussion}\label{Discussion}
\subsection{The setting of submodels interaction state}

In the design of fusion algorithms for multi-source navigation systems, adopting an Interactive Multiple Model (IMM) strategy is key to ensuring the collaborative functioning of submodels. This involves maintaining a unified reference benchmark for the common system states (i.e., public state variables) shared by multiple submodels, to ensure the overall consistency and efficacy of the algorithm. Even if the physical quantities and meanings are the same, inconsistency in reference benchmarks can easily lead to divergence in the system. For instance, consider setting up two submodels: Submodel A as IMU+GPS1 and Submodel B as IMU+GPS2, with each model's measurement prediction process updated independently. The common states are defined as \(\delta \bm{x}=  [\begin{array}{ccccc} \delta \boldsymbol{\theta}_I^{n \top} & \delta \bm{v}^{n \top} & \delta \bm{p}^{n \top} \end{array}]^{\top}\), including attitude error, velocity error, and position error. As filtering progresses, the error reference for the two models may shift, causing the error states \({^{(1)}\delta \bm{x}_k}\) and \({^{(2)}\delta \bm{x}_k}\) at the same moment \(k\) to lose correlation. In such cases, even if the main states of the two models are physically consistent, direct interactive weighted fusion may no longer be effective due to the uncorrelated nature of the error states.

This issue highlights that in designing fusion algorithms for multi-source navigation systems, it's essential not only to consider the physical consistency of the states of each submodel but also to ensure the consistency and correlation of the error states to maintain the effectiveness of the fusion algorithm. Therefore, the algorithm model in this paper separates prediction and measurement processes, ensuring that submodels share a common prediction process to maintain the same reference benchmark across submodels. This approach helps to align the error states across submodels, reducing the risk of divergence and enhancing the overall robustness and reliability of the fusion algorithm in multi-source navigation systems.

\subsection{The sensitivity of submodel weights}

In experiments, it is indeed possible to encounter a scenario where the fusion algorithm tends to favor submodels with larger measurement errors when the difference in measurement errors among submodels is not significant. In such cases, the fusion weights seem to adjust based on the error baseline of the submodels themselves, rather than their absolute measurement accuracy. For example, if at a certain moment submodel A has a higher fusion weight, the system, following the output of model A, may potentially consider model A to be more credible. From the perspective of model A, the measurement innovations and related covariances of model B will be compared with those of model A. Even if model B is relatively closer to the true value, since the reference frame is pulled towards model A, the measurement innovations and related covariances of model B might be relatively amplified, a situation colloquially referred to as a "trust crisis."

This "trust crisis" reflects the challenges faced by fusion algorithms in weight allocation and error evaluation, especially when the performance of different submodels is similar but not identical. This phenomenon is often caused by an inappropriate model transition probability matrix \(\pi\). As prior knowledge, \(\pi\) reflects the expected probability of transitions between different models to some extent, and the posterior probability of transitions between submodels is determined by both \(\pi\) and the likelihood of submodels \(^{(i)}L_k\). Therefore, to prevent the system from following measurements that cause a shift in the reference due to performance degradation, it is essential to set a \(\pi\) matrix that accurately reflects the credibility indexes of the system's sensors. Furthermore, increasing interaction and information sharing between models may also help improve the overall system's robustness and accuracy, thereby reducing the risks associated with a preference for a single model.
\section{Conclusion}\label{Conclusion}
This chapter initially establishes a simple and efficient dimensionless sensor credibility evaluation system. Based on credibility, the optimal multi-source sensor submodel combination architecture can be selected, providing key model prior knowledge for multi-source resilient fusion. Furthermore, this chapter proposes a multi-source resilient fusion framework based on Invariant Kalman Filtering, utilizing the defined sensor credibility indexes to guide the design of the model transition probability matrix. This approach reduces the sensitivity of submodel weights to fusion stability, addressing the issue of a lack of tuning basis for the model transition matrix. The algorithm proposed in this chapter has been validated with data from a redundant UAV flight platform. Through a comparison of pose solutions, the IMM-RIEKF proposed in this paper shows superior robustness and accuracy compared to single-model RIEKF. Moreover, it further simulates the resilient advantage of the IMM-RIEKF algorithm under multiple faults, such as heavy-tailed non-Gaussian noise and GPS jamming injections. This redundant interactive resilient fusion framework, without using any sensor fault diagnosis and isolation logic or setting any complex detection timings and thresholds, leverages the adaptive resilience advantage between models to achieve seamless switching between submodels. This enhances the adaptability of the state estimation system in complex environments, maximally utilizing the advantages of sensors to achieve a complementary enhancement of pose accuracy and robustness.



\ifCLASSOPTIONcaptionsoff
  \newpage
\fi



%
\bibliographystyle{IEEEtran}
\bibliography{refs}

%








\end{document}